\documentclass[3p,times]{elsarticle}

\usepackage{ecrc}
\usepackage{epstopdf}

\volume{00}

\firstpage{1}

\journalname{Nuclear Physics A}

\runauth{}

\jid{nupha}

\usepackage{amssymb}

\usepackage[figuresright]{rotating}

\begin{document}

\begin{frontmatter}



\dochead{}

\title{Measurements of di-jets in p+p and Au+Au in the STAR experiment.}


\author{Elena Bruna for the STAR Collaboration}

\address{Physics Department, Yale University, 272 Whitney Ave., New Haven, 06520-8120 CT, USA. }

\begin{abstract}

Jets are produced from hard scatterings in the early stages of heavy-ion 
collisions. It is expected that these high-p$_T$ partons
travel through the hot and dense medium before fragmenting.
Therefore they are expected to suffer energy loss in the QGP via gluon 
radiation or elastic collisions along their path.
Measurements from full jet reconstruction help in the understanding of energy loss and its effect§
on the jet structure and energy profile.
A data-driven characterization of the background in Au+Au is needed in 
order to compare the results to p+p.
The large coverage of the STAR detector along with an online 
trigger also allows for reconstructing di-jets.
Suitable selection of trigger jets allows for studying a subset of jets 
on the recoil side which are biased towards
higher energy loss because of a larger, on average,  in-medium path length traversed.
Trigger jets are also used to study jet-hadron correlations as an 
independent measurement to assess the effect of energy loss on the recoil 
side.
We present measurements of di-jets and jet-hadron correlations in the presence of reconstructed di-jets in Au+Au 
and p+p at 200 GeV in the STAR experiment.

\end{abstract}


\end{frontmatter}





\section{Introduction}
Full jet reconstruction reduces the geometrical biases~\cite{renk} inherent the study of jet quenching in single particle and di-hadron analyses, since it allows a more precise determination of the energy of the initial hard scattering of incident partons.
Di-jets can be exploited to probe the hot and dense medium produced in high energy nuclear collisions. Such measurements are possible in STAR thanks to its large acceptance coverage.
High momentum trigger jets, selected by the online trigger, tend to be biased towards partons that scatter near the surface of the medium~\cite{renk}.
The corresponding recoil jets on the away-side that are not emitted tangentially, are expected to travel over a large in-medium path-length due to the trigger bias.
Di-jet coincidence rates are compared in p+p and Au+Au in a scenario of broadening of jet structure in the medium. First jet-hadron correlation studies in the presence of a reconstructed di-jet are presented as a tool to assess  biases in jet-finding and to investigate the quenching scenario.
Modern jet-finding algorithms and data-driven correction schemes are used in this analysis.

\section{Data sets and analysis}\label{ana}
The STAR sub-detectors used for full jet reconstruction are the Time Projection Chamber (TPC) for charged particles and the Barrel Electromagnetic Calorimeter (BEMC) for the neutral energy. Both TPC and BEMC have full azimuthal coverage and pseudo-rapidity acceptance $|\eta|<1$.
Corrections for double-counting of electrons and charged hadronic energy deposition in the BEMC are applied. 
This analysis is based on p+p year 2006 and 0-20$\%$ most central Au+Au year 2007 events. Both data sets were selected with an online High-Tower (HT) trigger in the BEMC which requires the transverse energy above 5.4 GeV in a tower of size 0.05x0.05 in $\eta-\phi$.

The jet-finding utilizes recombination algorithms included in the FastJet package~\cite{fastjet1,fastjet2,fastjet3}. We use ``anti-$k_T$'' to reconstruct the signal jets and   ``$k_T$'' to estimate background per unit area.
The di-jets are defined by a ``trigger'' jet, that matches the online triggered tower in the BEMC, and a ``recoil'' jet, reconstructed on the away-side of the trigger jet (i.e. $|\Delta \phi| > 2.74$). Both trigger and recoil jets are reconstructed with a resolution parameter R=0.4 and are required to be fully included in the STAR acceptance (i.e. $|\eta_{jet}|<1-R$).

The background that is subtracted to get the reconstructed jet p$_{T}^{jet}$ is calculated by FastJet as the median $p_T$  per unit area ($\rho$) obtained from all the jets reconstructed with $k_T$. The trigger jets were reconstructed with a $p_T^{cut}=2$ GeV/c on tracks and neutral towers. This requirement together with the presence of the matching high calorimeter tower provides a selection bias of hard fragmenting jets produced in proximity of the surface of the medium, and/or less interacting. This selects a similar jet energy scale in p+p and Au+Au. A precise quantification of the trigger bias is currently under study. The requirement of $p_{T}^{jet}>20$ GeV/c for the reconstructed trigger jets minimizes the contamination by background jets. In order to study the bias introduced by a $p_T^{cut}$ on the recoil side, the recoil jets are reconstructed with two different choices of  $p_T^{cut}$, 0.2 and 2 GeV/c. 
The background spectrum consisting of uncorrelated background clusters and additional hard scatterings is estimated in this di-jet analysis by looking at the jet spectrum in a window centered at $\pm  \pi/2$ with respect to the trigger jet axis. 
The background spectrum is then subtracted from the recoil jet spectrum.
Region-to-region fluctuations are approximated by a Gaussian distribution with $\sigma=6.5$ GeV/c for $p_T^{cut}=0.2$ GeV/c~\cite{mateuszQM,elenaQM}. The recoil spectra with $p_T^{cut}=2$ GeV/c are unfolded assuming a Gaussian shape for the background fluctuation with width $\sigma=1.5$ GeV/c for R=0.4. Recent efforts are being made in STAR~\cite{peter}  to derive a data-driven form of the background fluctuations that accounts for the non-Gaussian tails. Preliminary studies suggest that the effect of non-Gaussian tails is small due to the flatness of the measured recoil spectrum.

The reconstructed jet momenta are corrected for the different tracking efficiencies in p+p and Au+Au (the difference is $\sim 10 \%$).

\section{Di-jet coincidence rate}\label{dijet}
\begin{figure}
\centering
\resizebox{0.49\textwidth}{!}{  \includegraphics{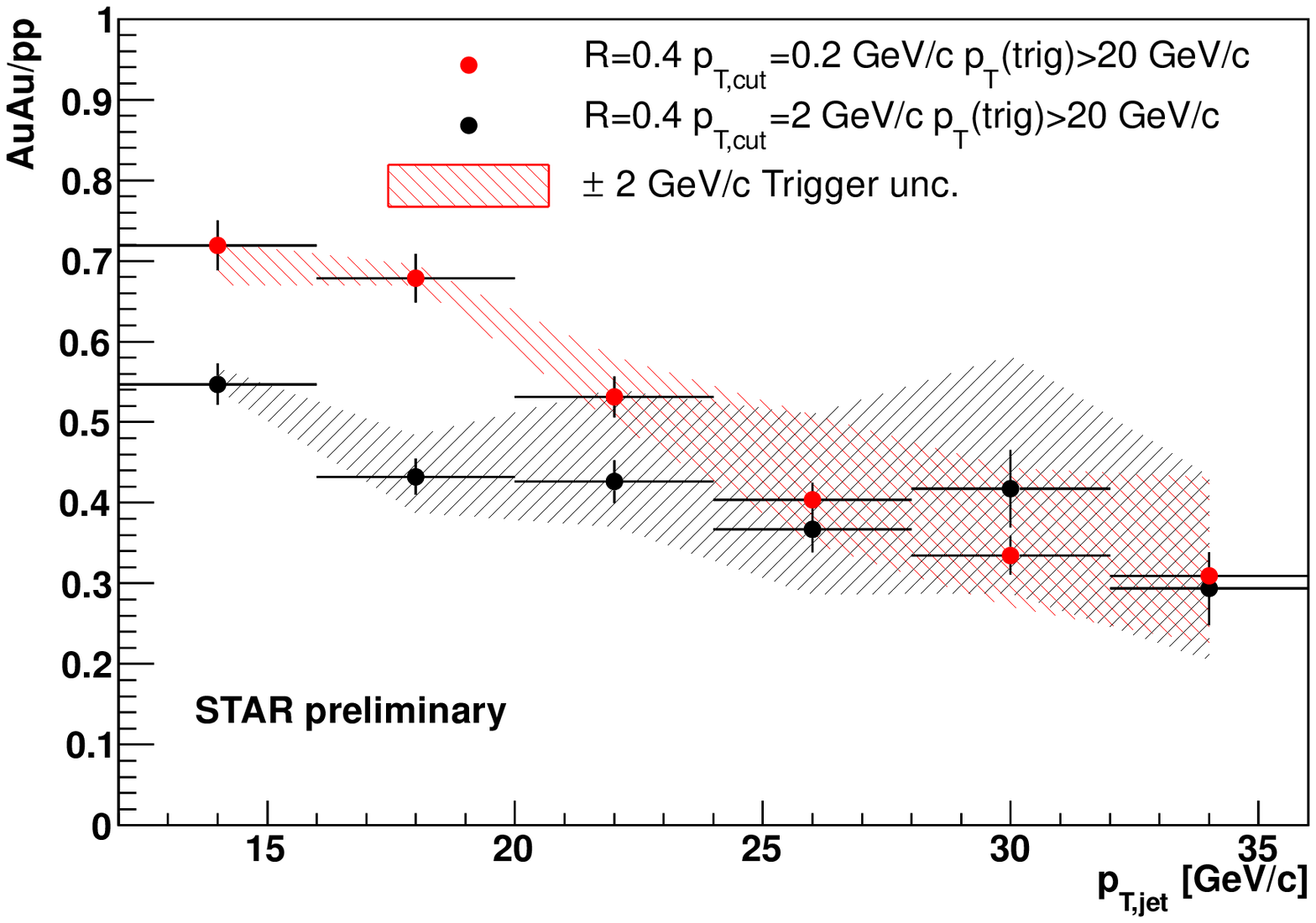}}
\resizebox{0.49\textwidth}{!}{  \includegraphics{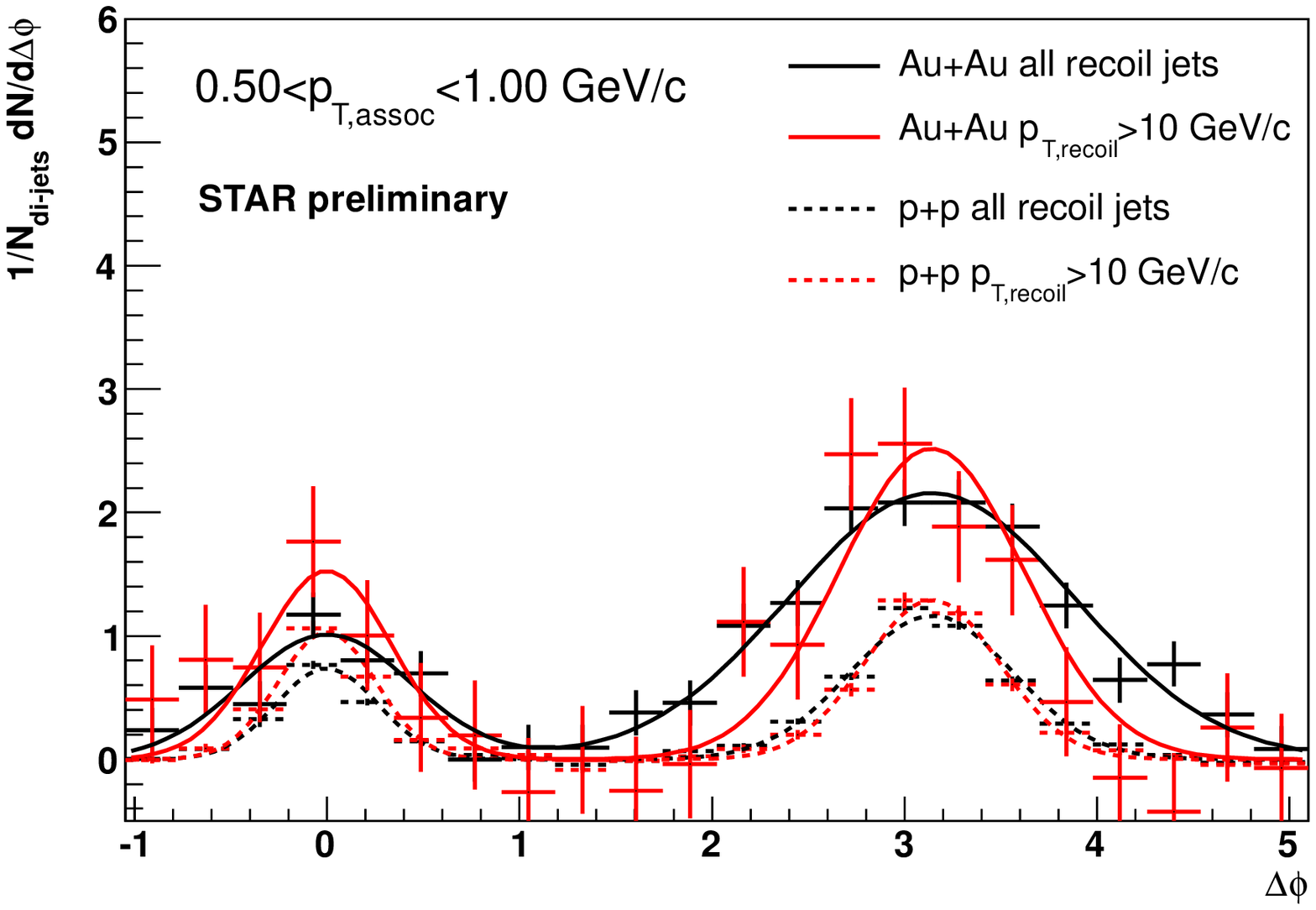}}

\caption{Left: Ratio of di-jet coincidence rates in Au+Au relative to p+p for two values of $p_T^{cut}$ on the recoil side, 0.2 and 2 GeV/c. The colored bands represent the uncertainty on the trigger jet energy. The trigger jet is selected with $p_{T,trigger}>20$ GeV/c. Right: Jet-hadron correlations for $0.5<p_{T,assoc}<1$ GeV/c in Au+Au (solid lines) and p+p (dashed), with no requirements on recoil jets (black) and with recoil jets selected above $p_{T,recoil}>10$ GeV/c (red). Same selection of trigger jets as left plot.}
\label{fig:dijet}
\end {figure}

The ratio of di-jet spectra in Au+Au relative to p+p is shown in Fig.~\ref{fig:dijet} (left), for two different values of $p_T^{cut}$ on tracks and neutral towers, $p_T^{cut}=0.2$ and 2 GeV/c.
A similar suppression is observed for the two different values of $p_T^{cut}$. 
A crucial step in this analysis is the control of the trigger energy in the two different collision systems. 
As mentioned in Sec.~\ref{ana}, a tight $p_T^{cut}$ is applied on the trigger jets to allow similar energy scales. 
However, the reconstructed trigger jet momentum might still be affected by upward background fluctuations of the order of $1-2$ GeV/c that artificially enhance the measured $p_{T}^{jet}$ compared to the same jet in p+p. 
On the other hand, given that trigger jets in Au+Au may actually interact in the medium (as observed from the near side of jet-hadron correlations~\cite{joern}), the measured $p_{T}^{jet}$ could correspond to a higher momentum jet in p+p. Both these effects are quantified by varying the trigger jet momentum in p+p by $\pm 2$ GeV/c, as indicated by the colored bands in  Fig.~\ref{fig:dijet} (left).

The observed suppression can be explained with a broadening of the recoil jet energy profile, due to energy loss to large angles. This would move part of the jet energy out of the jet area, shifting the reconstructed jet spectrum to lower $p_T$.
Assuming that broadening is responsible for a constant shift of the Au+Au recoil spectrum towards lower p$_T$ relative to p+p, we estimated that a shift of $7-8$ GeV/c for recoil jets with  $p_T^{cut}=0.2$ GeV/c and of $6-7$ GeV for $p_T^{cut}=2$ GeV/c  reproduces the measured ratios in left panel of Fig.~\ref{fig:dijet}. These preliminary results are consistent with jet-hadron correlation studies (see Sec.~\ref{JH}) and may suggest large angle radiation (i.e. due to broadening) of the recoil jets exposed to a maximum in-medium path.

As a consequence of the  high-$p_T$ hadron suppression related to the surface bias of high-$p_T$ particles~\cite{renk}, one could think that the recoil jets reconstructed with the tight requirement of  $p_T^{cut}=2$ GeV/c could be biased towards the surface of the medium in Au+Au, hence being less interacting and less subject to broadening. This issue can be investigated via jet-hadron correlations studied in presence of both trigger and recoil jets, as reported in the following section.

\section{Jet-hadron correlations with reconstructed di-jets}\label{JH}
Results reported above show that recoil jets reconstructed with $p_T^{cut}=2$ GeV/c are significantly suppressed, suggesting strong interactions of away-side jets with the medium.
The question we want to address is whether the reconstructed recoil jets are biased towards the surface of the medium, as we could expect because of the bias given by the  $p_T^{cut}$. 


This analysis is a first look at correlations of hadrons with respect to the trigger jets in events where di-jets are reconstructed. In this context, jet-hadron correlations are expected to show possible biases in the jet-finding via a direct measurement of the recoil side structure. Trigger and recoil jets are reconstructed with $p_T^{cut}=2$ GeV/c and R=0.4, both in p+p and Au+Au. The  trigger jets are required to have $p_{T}^{jet}>20$ GeV/c.

Figure~\ref{fig:dijet} (right) shows the azimuthal jet-hadron correlations for  $0.5<p_{T,assoc}<1$ GeV/c, in Au+Au (solid lines) and p+p (dashed), for the case where no requirement is applied on the recoil jet $p_T$ (black) and for recoil jets selected above $p_{T,recoil}>10$ GeV/c (red). The background level is estimated via a fit with a two-Gaussian function plus a $v_2$ contribution. The default jet $v_2$ value is the hadron $v_2\{2\}$ at 6 GeV/c. The associated hadron $v_2$ values are taken from~\cite{v2}. 
The results are not corrected for background jets. However,  about $1/30$ of the recoil jets with $p_{T}^{jet}>10$ GeV/c  reconstructed with  $p_T^{cut}=2$ GeV/c are expected to be background jets  (this is estimated via the jet spectrum at $\pi/2$ with respect to the trigger jet axis, see Sec.~\ref{ana}) . 
 
\begin{figure}
\centering

\resizebox{0.49\textwidth}{!}{  \includegraphics{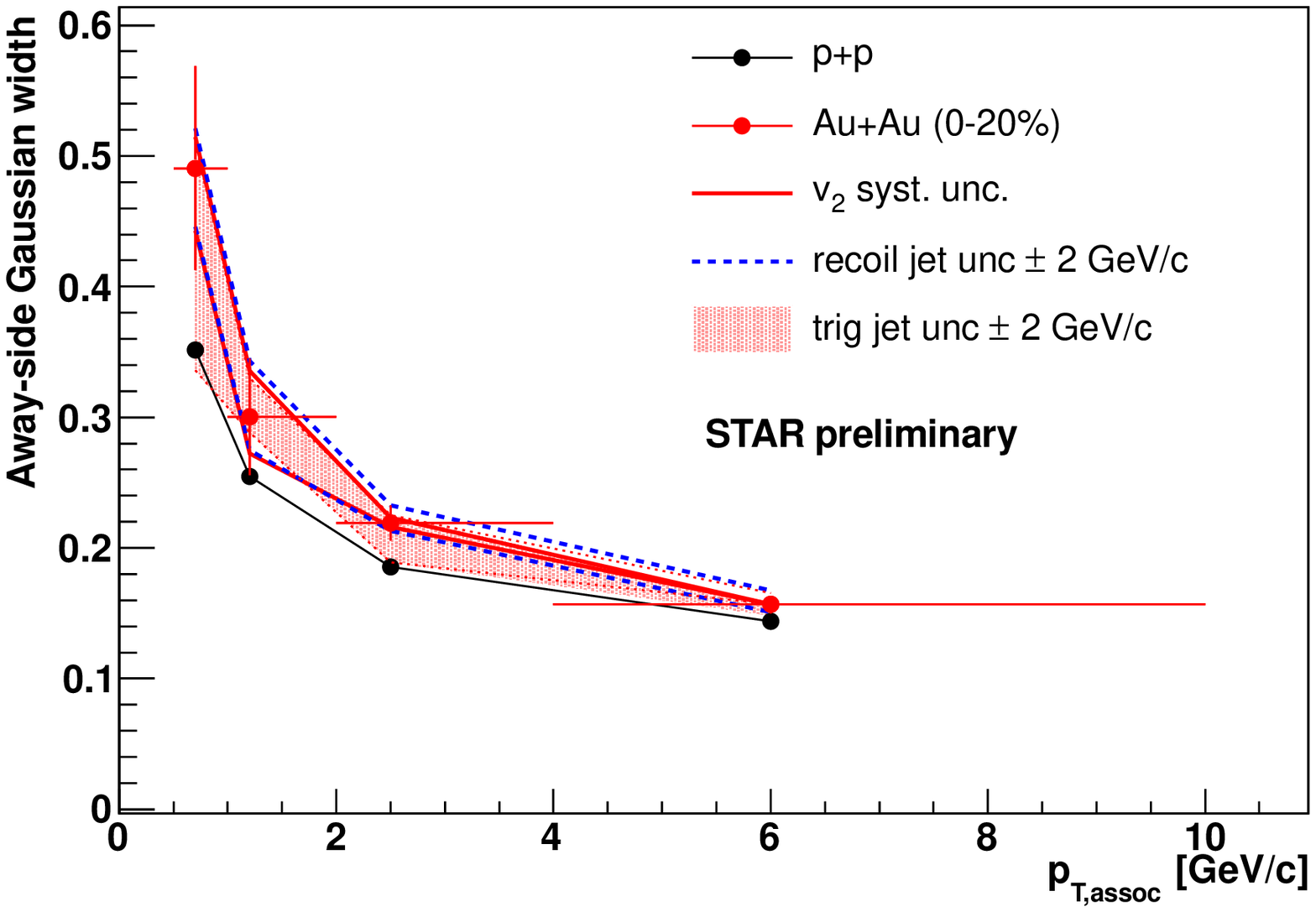}}
\resizebox{0.49\textwidth}{!}{  \includegraphics{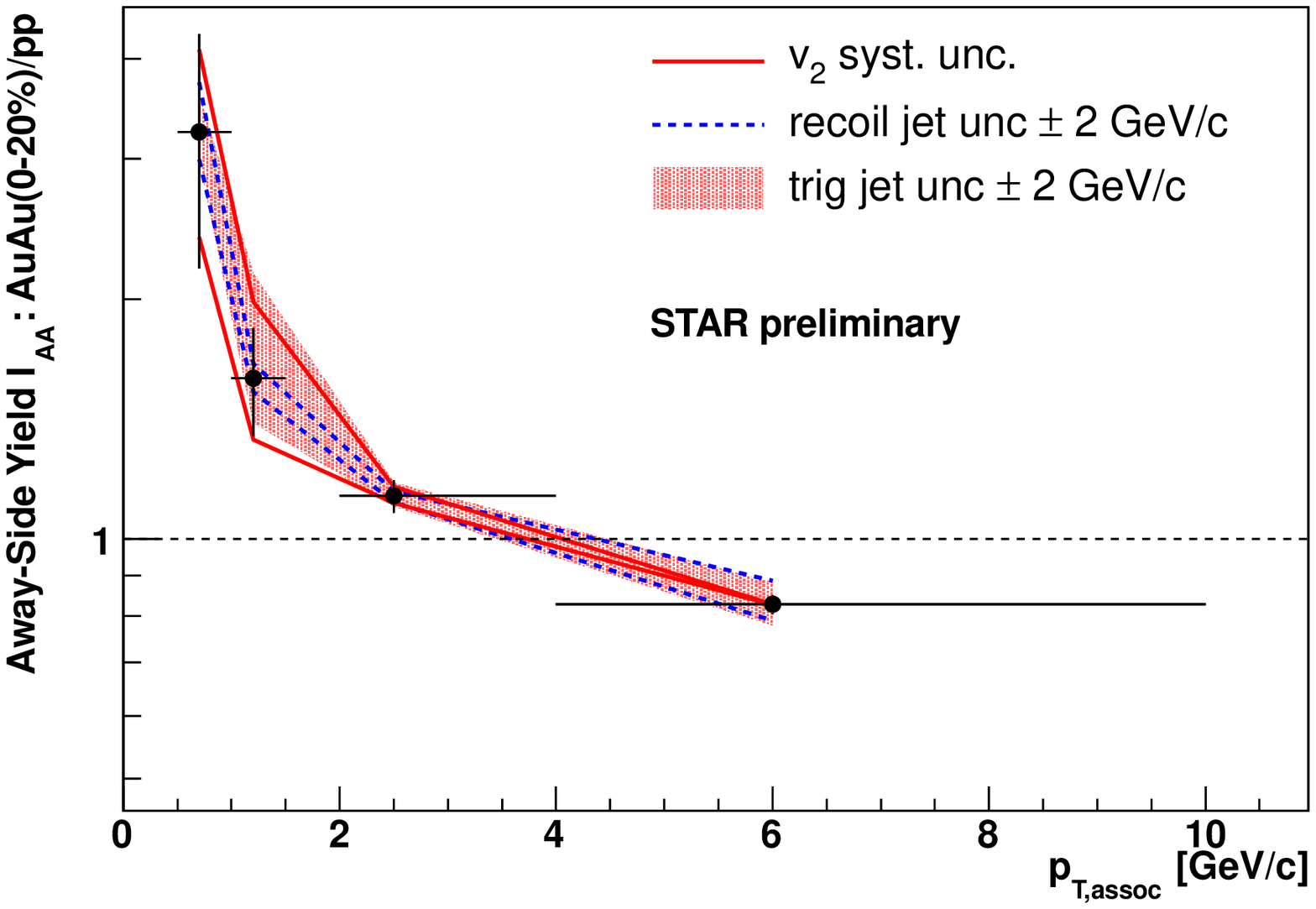}}
\caption{Left: Gaussian width of the away-side peak of jet-hadron correlations as a function of $p_{T,assoc}$, for Au+Au (red) and p+p (black), in the presence of a recoil jet having $p_{T}^{jet}>10$ GeV/c and $p_T^{cut}=2$ GeV/c. The systematic errors are described in the text. The trigger jet is selected with $p_{T,trigger}>20$ GeV/c. 
Right: $I_{AA}$ of the away-side peak of jet-hadron correlations as a function of $p_{T,assoc}$, in the presence of a recoil jet having $p_{T}^{jet}>10$ GeV/c and $p_T^{cut}=2$ GeV/c. The systematic errors are described in the text. Same selection of trigger jets as left plot.}
\label{fig:jh}

\end {figure}

A broadening of the the away-side and an enhancement of low-$p_T$ particles in Au+Au relative to p+p can be observed in Fig.~\ref{fig:dijet} (right).
This effect is observed in both cases with and without requirements on the $p_{T}$ of the recoil jet. 
The away-side Gaussian width is reported in Fig.~\ref{fig:jh} (left) as a function of $p_{T,assoc}$ for the case where recoil jets are selected with $p_{T}^{jet}>10$ GeV/c and $p_T^{cut}=2$ GeV/c.
At high  $p_{T,assoc}$ the Gaussian width of the away-side is similar in Au+Au and p+p, indicating only a minor broadening of the hard sector of the jet fragmentation. 
The systematic uncertainties represent the $\pm 2$ GeV/c uncertainty on the energy scale for both trigger (red shaded area) and recoil jets (blue dashed lines) that is expected due to energy loss and background fluctuations above $p_T^{cut}=2$ GeV/c. The uncertainty band due to the choice of the jet $v_2$ is also reported (red solid lines). The lower limit is the case where no jet $v_2$ is assumed. The upper limit is the case where the jet $v_2$ is 50$\%$  higher than the hadron $v_2\{2\}$ at 6 GeV/c.
The softening of recoil jets is clearly visible in Fig.~\ref{fig:jh} (right),  where the ratio of the away-side peak in Au+Au relative to p+p (I$_{AA}$) is shown for different $p_{T,assoc}$ bins.
The decrease of high-$p_T$ associated particles and the corresponding enhancement of the low-$p_T$ particles in Au+Au with respect to p+p suggests that also recoil jets reconstructed with a large bias due to the tight $p_T^{cut}$ are subject to the effects of energy loss.

 Preliminary results on jet-hadron correlations~\cite{joern} show that the average energy outside R=0.4 is of the order of 6-9 GeV. This could indicate, in agreement with Sec.~\ref{dijet}, that the measured suppression of the di-jet coincidence can be explained in terms of a broadening scenario due to energy loss in Au+Au that leads to an underestimation of the jet energy.

\section{Summary}
The measured suppression of the di-jet coincidence rate in Au+Au relative to p+p is in qualitative agreement with a scenario of jet broadening outside of R=0.4. This observation can be interpreted as the effect of the large in-medium path-lenght that non-tangentially emitted recoil jets must traverse, given the tight requirement on the trigger jets that bias them towards the surface.
Jet-hadron correlations performed in the presence of reconstructed di-jets are used both to address possible biases in the jet-finding and to quantify the missing out-of-cone jet energy due to quenching effects. 
The observed broadening and softening of jet fragmentation for the highly biased population of recoil jets reconstructed  with  $p_T^{cut}=2$ GeV/c suggests that the jet finding is not biased towards only surface or non-interacting jets.

Completion of the corrections, assessment of the trigger bias and of the effect of $v_2$ in the background subtraction will provide a quantitative estimate of the out-of-cone energy loss via jet-hadron correlations.
The use of quenching models will be useful to compare the measured out-of-cone energy with the theoretical predictions.

\section{Acknowledgments}
The author wishes to thank the Bulldog computing facility at Yale University.


\begin{thebibliography}{00}
\bibitem{renk} T. Renk, K. Eskola, Phys. Rev. C75 (2007) 054910, arXiv:hep-ph/0610059.
\bibitem{fastjet1} M. Cacciari, G. Salam, Phys. Lett. B64, 57 (2006), arXiv:hep-ph/0512210v2. 
\bibitem{fastjet2} M. Cacciari, G. Salam, G. Soyez, JHEP 0804, 005 (2008), arXiv:0802.1188. 
\bibitem{fastjet3} M. Cacciari, G. Salam, G. Soyez, JHEP 0804, 063 (2008), arXiv:0802.1189v2. 
\bibitem{mateuszQM} M. Ploskon for the STAR Collaboration, Quark Matter 2009 Proceedings, Nucl. Phys. A830 (2009) 255c, arXiv:0908.1799.
\bibitem{elenaQM} E. Bruna for the STAR Collaboration, Quark Matter 2009 Proceedings, Nucl. Phys. A830 (2009) 267c, arXiv:0907.4788.
\bibitem{peter} P. Jacobs for the STAR Collaboration, these proceedings.
\bibitem{joern} J. Putschke for the STAR Collaboration, these proceedings.
\bibitem{v2} I. Abelev et al. (STAR), Phys. Rev. C80, 064912 (2009), arXiv:0909.0191
\end{thebibliography}
\end{document}